\documentclass[12pt,preprint]{aastex}

\tighten

\shorttitle{False Lensing Events in M22}
\shortauthors{Sahu, Anderson, \& King}

\begin{document}

\title{A RE-EXAMINATION  OF THE ``PLANETARY'' LENSING EVENTS IN
M22\footnote{Based on observations with the NASA/ESA {\it Hubble Space
Telescope}, obtained at the Space Telescope Science Institute, which is
operated by AURA, Inc., under NASA contract NAS 5-26555.}}

\author{Kailash C. Sahu}
\affil{Space Telescope Science Institute, 3700 San Martin Drive,
Baltimore, MD 21218\\
E-mail: ksahu@stsci.edu}

\author{Jay Anderson and Ivan R. King}
\affil{Astronomy Department, University of California, Berkeley, CA
94720\\
E-mail: jay@cusp.berkeley.edu, king@glob.berkeley.edu}

\begin{abstract}
We have carried out further analysis of the tentative, short-term
brightenings reported by Sahu et al.\ (2001), which were suggested to be
possible lensings of Galactic-bulge stars by free-floating planets in
the globular cluster M22.  Closer examination shows that---unlikely as
it may seem---small, point-like cosmic rays had hit very close to the
same star in both of a pair of cosmic-ray-split images, which
cause the apparent brightenings of stars at the times and locations
reported. We show that the observed number of double hits is consistent
with the frequency of cosmic rays in WFPC2 images, given the number of
stars and epochs observed.  Finally, we point to ways in which cosmic
rays can be more directly distinguished.
\end{abstract}


\clearpage

\section{Background}
Sahu et al.\ (2001) have recently reported observations of microlensing
of stars in the Galactic bulge by stars of the globular cluster M22.
They report one major event, with a characteristic time of $\sim$18 days
and a brightening by 3 magnitudes, and six brightenings of 0.3 to 0.8
magnitudes, each seen as similar brightenings in both images of a pair
of images taken 6 minutes apart.  We discuss here only the short-term
events.  We demonstrate that each pair of brightenings is caused by two
separate cosmic ray hits, one in each image of the CR-split, that happen
to occur near the same star.  In \S\ 2 we describe the observations and
our re-analysis; in \S\ 3 we present the evidence that the short-term
events are caused by cosmic rays; in \S\ 4 we discuss the prevalence of
CRs, and in \S\ 5 we suggest effective ways of reliably avoiding CR
contamination.

Since there are already at least five papers (Gaudi 2001, de la Fuente
Marcos \& de la Fuente Marcos 2001, Hurley \& Shara 2001, Fregeau et
al.\ 2001, Soker, Rappaport, \& Fregau 2001) that discuss the
short-term events, we feel that it is urgent to report our new finding,
based on further analysis.

\section{The data and initial analysis} 

The observations of Sahu et al.\ were made with the WFPC2 camera of the
{\it Hubble Space Telescope} ({\it HST}) between 22 Feb 1999 and 15 Jun
1999 (GO 7615), to determine the frequency and nature of lensing of
bulge stars by those of M22, the details of which are given by Sahu et
al (2001). Upon seeing the results of Sahu et al., the two of us who
had not been part of the original team (J.A.\ and I.R.K.) were eager to
apply our astrometric techniques (Anderson \& King 2000, King et al.\
1998, Bedin et al.\ 2001) for the measurement of proper motions, to
determine whether the lensed stars were in fact bulge members or if
they were cluster members, which would imply a different lensing
object, somewhere between the cluster and us.  There exist in the {\it
HST} Archive images taken in 1994 and 1995, which provided an excellent
baseline for such a separation.

Close examination of the images for the purpose of identifying the
``brightened'' stars raised questions, however.  In each pair  of
CR-split images, the brightness pattern of the pixels around the
``brightened'' star differed between the two images, much more than did
the pixels around other stars of comparable magnitude, which made a
convenient reference standard.  This called for further investigation.
We immediately contacted the P.I.\ of the original paper, and the
present Letter is the result.  (We regret that it has been delayed by
trips for observing and summer meetings.)

\section{Detailed analysis of the short-term events} In
Table~\ref{events} we give for each of the events the name, RA, Dec,
and date of brightening, from Sahu et al.\ (2001), along with the date,
the Dataset Name of the image (in brackets, the number of the chip in
which the event occurred), and the pixel coordinates of the star that
was affected.  (In two cases, the indicated region appeared in more
than one pointing, as is evident in Table 1.  The second pair was taken
about 9 minutes after the first pair.) We then compared the images in
question with images taken before and after, in order to identify
easily the star that had changed. Note that there was a transcription
error in the position and date of event B as given by Sahu et al.; it
has been corrected here.  To explain that the ``brightenings'' were
caused by point-like cosmic-ray hits close to the same star in both
images, we now show two of the events in detail, as images and in
numerical form.

\begin{figure}
\plotone{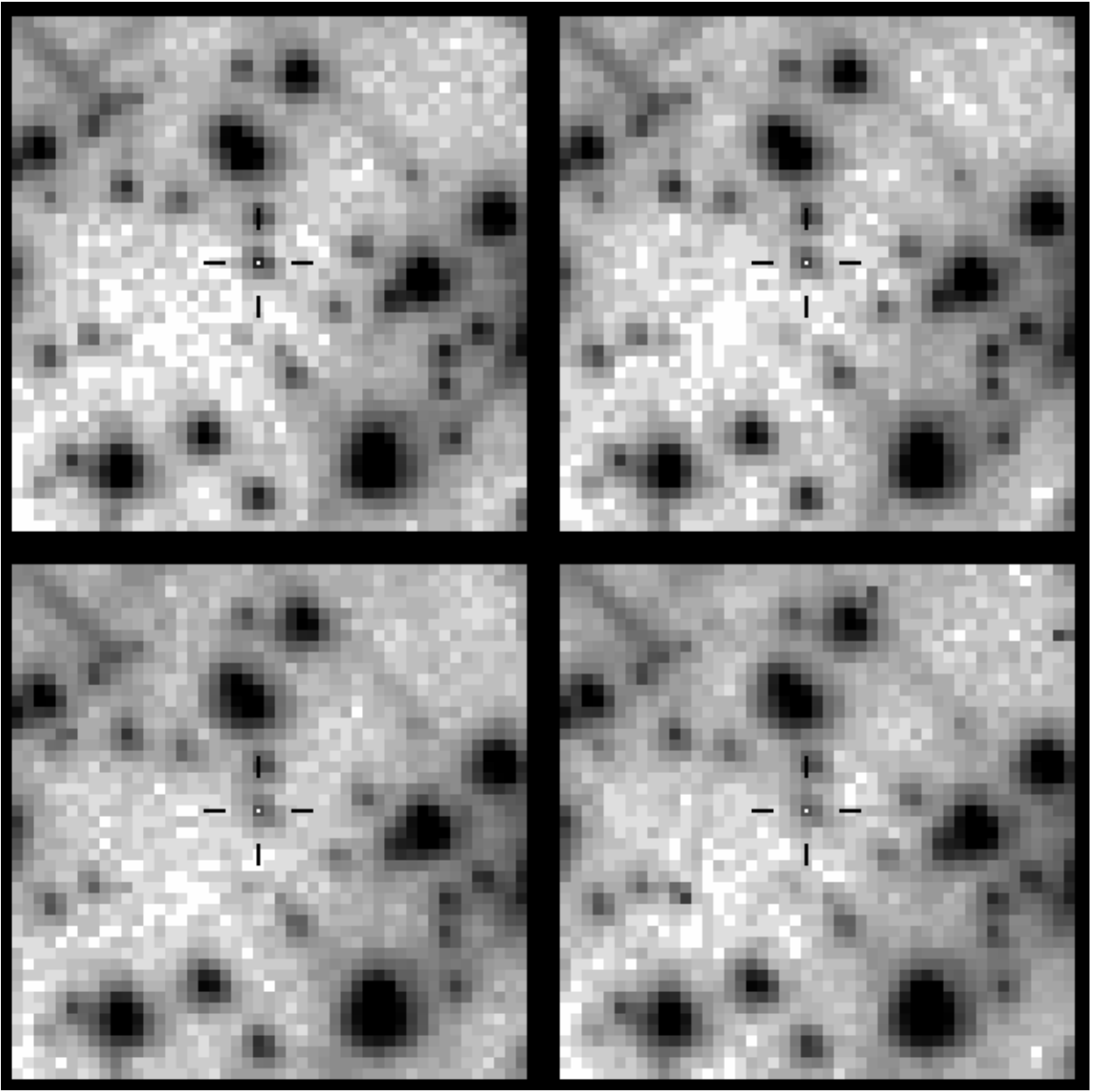}
\caption{The region surrounding the location of star D.   The white
dots identify the pixel within which the centroid of the star should
fall, as transformed from other images that show no brightening. 
See text for the source of the 4 panels.}

\label{figD}
\end{figure}

Figure~\ref{figD} shows event D, reproducing a small section of the
images at a large enough scale to allow examination of individual
pixels.  In the top two panels are the two images of the CR split in
which the event occurred; the lower panels show the image pair at
another time when the placement with respect to pixel boundaries
happened to be the same but there was no brightening.  The white dots
identify the pixel within which the centroid of the star should fall, as
transformed from other images that show no brightening.  

We have chosen this case for illustration because the two
``brightened'' images of star D do indeed resemble each other and
appear to be centered the same; on the basis of such a visual
inspection an observer might be inclined to accept them.  Despite
these similarities, both brightenings are best explained by cosmic
rays.

Table~\ref{tabD} shows, in its upper half, a comparison of pixel values
between the two images of the star. Each array gives the inner
$5\times5$ pixels of the star, in each of the two CR-split images, and
then the difference between them.  In the lower set of arrays are the
corresponding numbers for a star of similar brightness, in the same two
images.  In all cases pixels are labeled by row and column numbers.

The arrays labeled ``Difference" are quite revealing.  The comparison
star has very small differences, but the ``brightened" star shows large
differences between the two images, something that would not be the case
for real star images.

Furthermore, a careful measurement of the centroids of the star in the
two ``brightened'' images shows that they too differ by an amount which
is much larger than the centroid shifts exhibited by other stars.   We
will demonstrate this in detail in \S\ 5.  

In Figure \ref{figE} we show event E, which is more typical of the other
events.  Here the resemblance of the two images to each other is much
less.  As would be expected, the numbers in Table \ref{tabE} look worse,
and, as we shall see, the positional discrepancies are
larger. Nevertheless, even in this case each of the individual images
looks like that of a star.  It is only when we compare them with each
other, and with other pairs of star images, and make the additional
tests that we describe, that it becomes clear that the event is not a
real brightening.

The other events are qualitatively similar to event E, and we omit
displaying them here.

One might conceivably imagine other effects---all of them extremely
unlikely---that could cause positional shifts or differences in PSF
shape.  But even though we are convinced that the "brightenings" are
caused by cosmic rays, in a sense it is not essential to know whether it
was that or some more esoteric cause; what really matters is that these
differences and shifts exclude microlensing as the cause.  It is
conceivable, for example, that the source is blended (either because it
is part of a binary, or because another star happens to lie very close
to the same location); microlensing could then cause a shift in the
centroid (Dominik \& Sahu, 2000).  However, apart from the fact that two
such blended components would most likely be resolved in our case, the
blending cannot explain the {\it relative} centroid shift between the
two CR-split images.  An additional argument against microlensing as the
cause of apparent brightening is the fact that for event B and for event
F there was a second pair of exposures taken only 9 minutes later, which
do not show any brightening.

For all the reasons that we have explained, we believe that microlensing
is excluded as a possible cause of the ``brightenings.''  Of the other
possible causes, cosmic rays seem the only likely one---especially since
we are about to show that it is statistically probable that this many
paired CR hits would occur.

\section{The likelihood of a double CR event}
At first it might seem extremely unlikely that in a cosmic-ray split
pair, both images of the same star would be affected by cosmic rays in a
similar way.  We have noted from direct examination, however, that this
set of images has a cosmic-ray hit of about the strength observed (20 to
75 DN) in about about one pixel out of every 2000.  Since a hit by a
point-like cosmic ray in any one of half a dozen pixels around the
center of the star will produce the effect that we are discussing, both
images of a pair will be hit about one time in $10^5$.  If we note that
there were dozens of observations, with (conservatively) 30,000 stars in
each, this makes more than a million pairs of star images.  Thus 10 or
so double hits are to be expected, so that 6 is not a surprising number
after all.

The above is a very approximate calculation, especially in judging which
CR hits are ``of a similar strength'' and also pass the
visual-inspection criteria described by Sahu et al., but it does make
the point that an appreciable number of double hits is not unlikely.

\section{Reliable rejection of cosmic rays}
There is a lesson to be learned from our unhappy experience.  Cosmic
rays are everywhere and can indeed ``strike twice in the same place,''
in a way that can pass visual comparisons.  How then are we to avoid
them?  We recommend both of the tests that we have applied here.
Differencing entire images can be cumbersome, however---not in taking
the differences but in interpreting them.  Bright stars generate large
Poisson fluctuations in their differences, so that a difference map has
to be compared with the original image in order to interpret it.  The
power of the differencing method is in testing individual objects, as we
have shown here.

For general screening against cosmic rays, we recommend careful
measurement and comparison of positions.  Our measurements of stellar
flux, for example, involve a simultaneous measurement of position
(Anderson \& King 2000).  We can compare this with the star's average
position (from all the observations), transformed into the coordinate
frame of this image.

\begin{figure}
\plotone{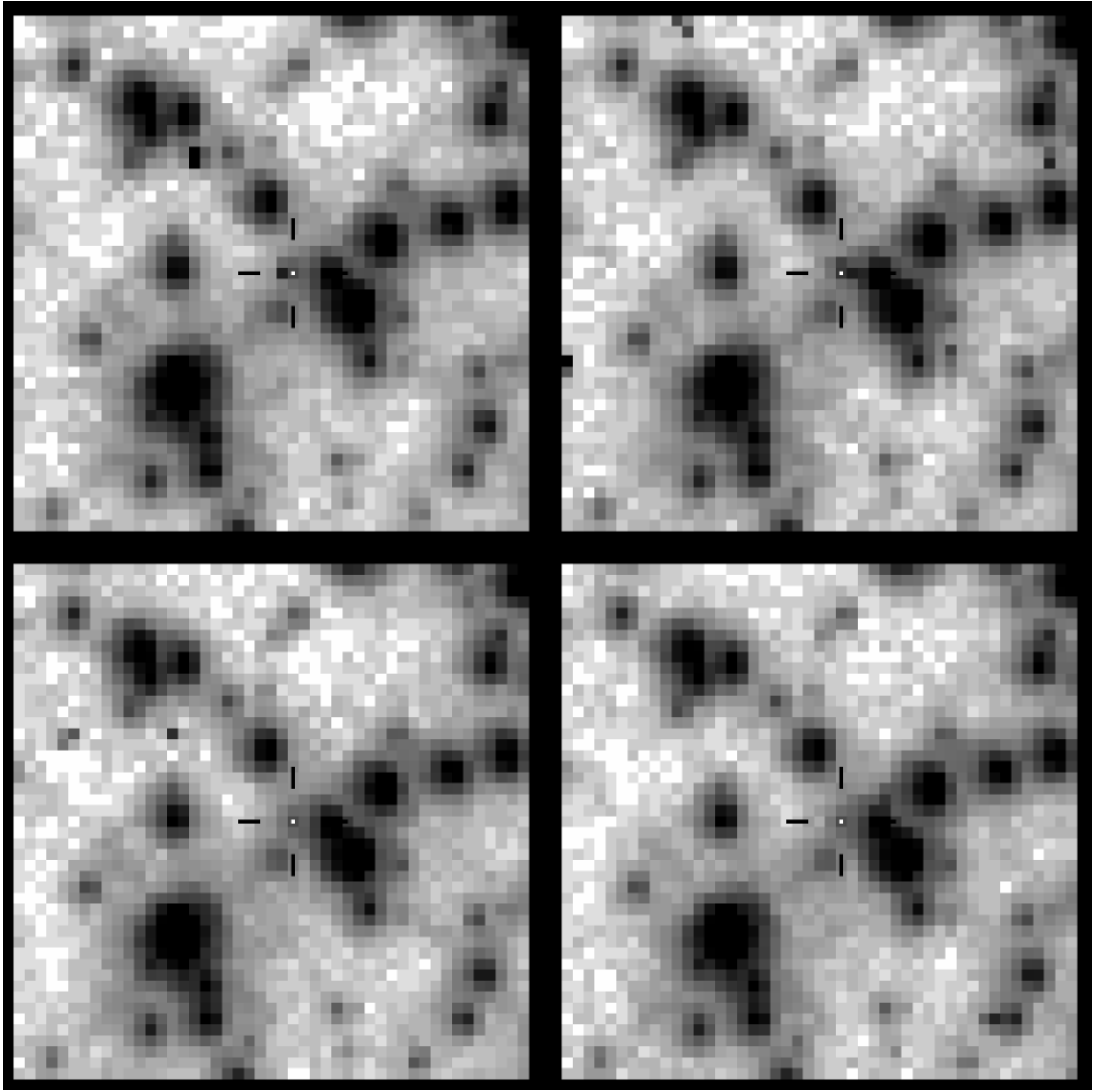}
\caption{The region surrounding the location of star E, displayed in the
same way as the previous figure.}
\label{figE}
\end{figure}

\begin{figure}
\plotone{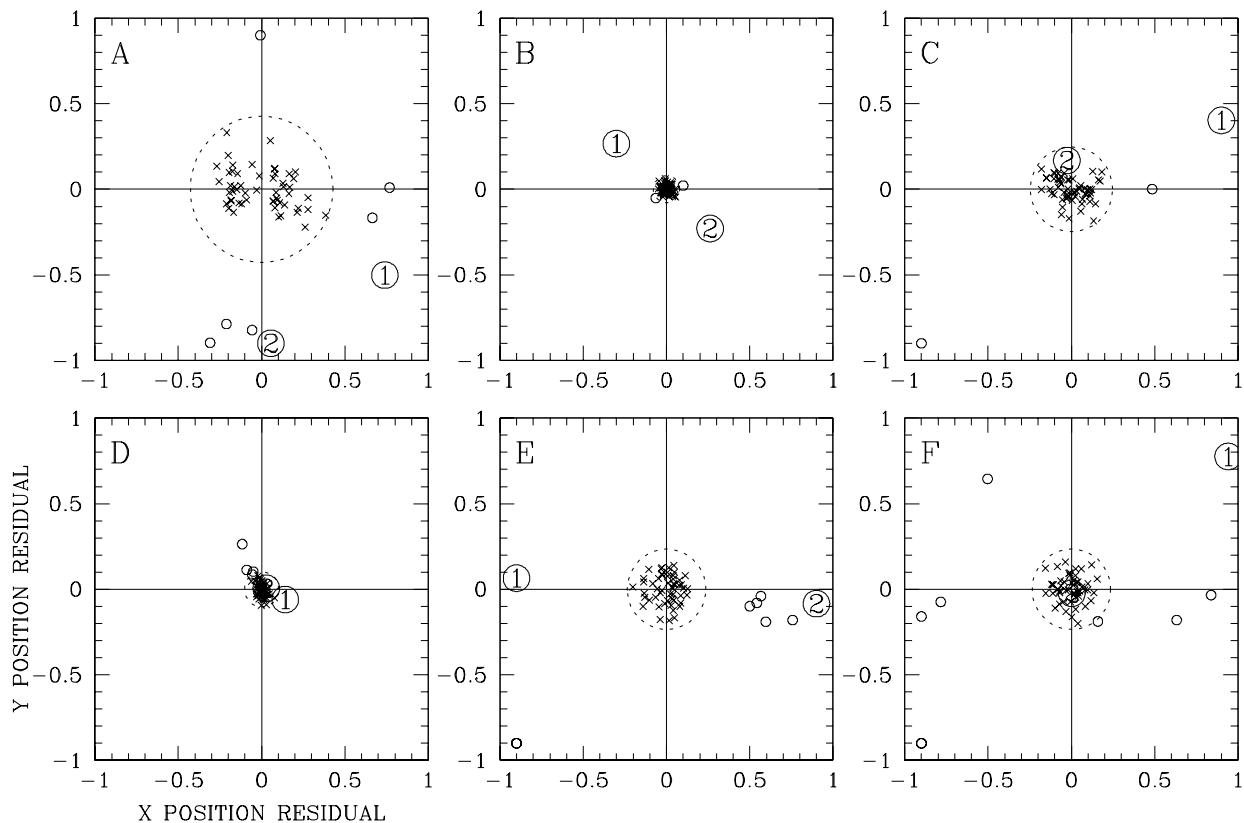}
\caption{The position residuals for all measurements
         of each of the event stars; residuals that lie beyond the bounds
         of the plot are shown at the edge.  The ``brightened'' observations
         are shown as circled numbers.  The dotted circle in each plot
has a radius of $3\sigma$; in a 2-D Gaussian 1.1\% of points should lie
outside this circle.  The open circles represent those
         measurements whose flux should be considered suspect.  
 }
\label{posfig}
\end{figure}

Figure~\ref{posfig} shows the position residuals of the $2\times31$
F814W observations of each of the 6 stars involved in these events.  (In
the case of events B and F, each of which fell in two pointings, we used
only the PC, which was the chip in which the events were seen.)  The
3-$\sigma$ error circle is shown in each case.  Any observation whose
measured position falls outside of this circle is unlikely---for some
reason, possibly a cosmic ray or a bad pixel---to have an accurately
measured flux.  We note that many of these particular stars are faint
and superposed on a mottled background due to the halos of bright stars,
so that it is natural to have some unusually large position errors.
Star E, as seen in Fig.\ \ref{figE}, is a good example; the size of its
error circle in Fig.\ \ref{posfig} is a good indication too.  Note also
that the asymmetries that we noted earlier in Fig.\ \ref{figE}
correspond to large position discrepancies.

As just explained, ``outlier'' measurements are not uncommon,
particularly for stars as faint as these.  In calculating mean positions
we therefore used what can be called ``iterative sigma-clipping.''
First we calculate a mean and a sigma from all the observations (the
latter from a percentile-based algorithm, so as to avoid giving undue
influence to the outliers).  Then we reject all measurements that are
outside a circle whose radius is 3 times the single-coordinate sigma.
This process is iterated until it converges (usually after no more than
one or two iterations).  Thus the circle symbols in Fig.\ 3 denote
points that were not used in calculating the means and sigmas.

\section{Conclusion}  We conclude that the 6 minor events found
in the GO 7615 observations of M22 can best be explained by coincident
cosmic rays rather than by gravitational lensing.  Indeed, Sahu et al.\
(2001) stated that ``The interpretation of these events as microlensing
is necessarily tentative.''  That caution was even more appropriate than
it seemed at the time.
  
Although these apparent brightenings are not caused by microlensing, we
should note that microlensing remains a sensitive technique to detect
the presence of small-mass objects in a globular cluster.

\acknowledgments We acknowledge useful consultations with Ron
Gilliland, Stefano Casertano, Michael Albrow, Mario Livio, and Nino 
Panagia. Support for this work was provided by NASA through Grant GO
7615 from the Space Telescope Science Institute.  J.A.\ and I.R.K.\
were supported by Grant GO 8153 from the Space Telescope Science
Institute.

\bigskip
{\parindent=0truemm
{\bf References:}

Anderson, J., \& King, I.\ R. 2000, PASP, 112, 1360

Bedin, L.\ R., Anderson, J., King, I.\ R., \& Piotto, G.  2001, ApJ,
560, L75

de la Fuente Marcos, R., \& de la Fuente Marcos, C.  2001, A\&A, 379, 872

Dominik, M., \& Sahu, K.\ C. 2000, ApJ, 534, 213

Fregeau, J.\ M., Joshi, K.\ J., Portegies Zwart, S.\ F., \& Rasio, F.\
A.  2001, ApJ, submitted, astro-ph/0111057

Gaudi, B.\ S. 2001, ApJ, in press; astro-ph/0108301

Hurley, J. R., \& Shara, M. 2001, ApJ, submitted; astro-ph/0108350

King, I.\ R., Anderson, J., Cool, A.\ M., \& Piotto, G. 1998, ApJ, 492, L37

Sahu, K.\ C., Casertano, S., Livio, M., Gilliland, R.\ L., Panagia, N.,
Albrow, M.\ D., \& Potter, 
\hspace*{0.3in} M.  2001, Nature, 411, 1022

Soker N., Rappaport, S., \& Fregau, J. 2001, ApJ, 563, L87
}

\clearpage

\begin{table*}
\caption{Identification of the short-term events.  The day of the year
is given for purpose of coordination with the times in Fig.\ 2 of Sahu
et al.\ (2001).}
\begin{center}
\begin{tabular}{ccccrcc}

Name  & RA(2000) & Dec(2000)     & Date  & DOY & Dataset Name  & location  \\
\hline
\hline
A & 18:36:28.63  & $-$23:53:55.6 & Apr01 &  91 & u5331901r[4]  & (359,225) \\
\hline
B & 18:36:26.08  & $-$23:53:36.6 & Mar14 &  73 & u5331001r[1]  & (629,235) \\
  &              &             &         &     & u5331003r[3]  & (567,520) \\ 
\hline
C & 18:36:31.24  & $-$23:53:12.5 & Mar06 &  65 & u5330501r[2]  & (415,350) \\
\hline
D & 18:36:28.14  & $-$23:52:57.5 & Apr30 & 120 & u5332901r[2]  & (170,245) \\
\hline
E & 18:36:24.09  & $-$23:55:51.9 & Apr16 & 107 & u5332405r[4]  & (140,434) \\
\hline
F & 18:36:26.07  & $-$23:53:36.6 & Mar14 &  73 & u5331001r[1]  & (625,226) \\
  &              &             &         &     & u5331003r[3]  & (569,523) \\
\hline
\label{events}
\end{tabular}
\end{center}
\end{table*}

\clearpage

\begin{table*}
\caption{Event D}
\begin{center}
\begin{tabular}{crrrrrrrrrrrrrrrr}
\hline
D  & \multicolumn{5}{c}{ CR-split 1 }
   & \multicolumn{5}{c}{ CR-split 2 }
   & \multicolumn{5}{c}{ Difference } \\
\hline
   & \multicolumn{5}{c}{ u5332901r\_c0f.fits[2] }
   & \multicolumn{5}{c}{ u5332902r\_c0f.fits[2] } & \\
\#1&~~167& 168& 169& 170& 171&   ~~167& 168& 169& 170& 171&   ~~167& 168& 169& 170& 171\\
\hline
248&   10&  11&  12&  10&   6&       8&  13&  12&  10&   6&      -2&   2&   0&   0&   0\\
247&    6&  17&  31&  14&   7&       7&  19&  30&  16&   8&       1&   2&  -1&   2&   1\\
246&   10&  18& 129&  42&  12&       7&  18&  72&  24&  10&      -3&   0& -57& -18&  -2\\
245&    7&  13&  20&  22&  11&       7&  13&  16&  19&  11&       0&   0&  -4&  -3&   0\\
244&    5&   8&   8&   9&   8&       7&  10&   6&   8&   8&       2&   2&  -2&  -1&   0\\
\hline
\hline
\#2&  167& 168& 169& 170& 171&     167& 168& 169& 170& 171&      167& 168& 169& 170& 171\\
\hline
252&   12&  13&  16&  10&   9&       9&  17&  18&  12&  10&      -3&   4&   2&   2&   1\\
251&   11&  17&  48&  22&  10&       9&  19&  53&  26&   8&      -2&   2&   5&   4&  -2\\
250&   11&  20&  83&  36&  12&      11&  21&  79&  37&  11&       0&   1&  -4&   1&  -1\\
249&    9&  14&  22&  16&   8&      11&  14&  19&  19&  10&       2&   0&  -3&   3&   2\\
248&   10&  11&  12&  10&   6&       8&  13&  12&  10&   6&      -2&   2&   0&   0&   0\\
\hline
\label{tabD}
\end{tabular}
\end{center}
\end{table*}

\clearpage

\begin{table*}
\caption{Event E}
\begin{center}
\begin{tabular}{crrrrrrrrrrrrrrrr}
\hline
E & \multicolumn{5}{c}{ CR-split 1 }
  & \multicolumn{5}{c}{ CR-split 2 }
  & \multicolumn{5}{c}{ Difference } \\
\hline
  & \multicolumn{5}{c}{ u5332405r\_c0f.fits[4] }
  & \multicolumn{5}{c}{ u5332406r\_c0f.fits[4] } & \\
\#1 &~~139& 140& 141& 142& 143&~~139& 140& 141& 142& 143&~~139& 140& 141& 142& 143\\
\hline
436 &   12&  15&  18&  22&  30&   14&  15&  16&  24&  31&    2&   0&  -2&   2&   1\\
435 &   11&  31&  36&  31&  47&   16&  29&  68&  45&  45&    5&  -2&  32&  14&  -2\\
434 &   14& 137&  58&  39&  68&   13&  25&  85& 139& 116&   -1&-112&  27& 100&  48\\
433 &   15&  22&  29&  33&  64&   18&  20&  28&  37&  92&    3&  -2&  -1&   4&  28\\
432 &   19&  25&  23&  29&  52&   20&  25&  22&  27&  49&    1&   0&  -1&  -2&  -3\\
\hline
\hline
\#2 &  128& 129& 130& 131& 132&  128& 129& 130& 131& 132&  128& 129& 130& 131& 132\\
\hline
436 &   17&  35&  57&  57&  16&   19&  33&  62&  57&  17&    2&  -2&   5&   0&   1\\
435 &   20&  40& 183& 174&  26&   19&  46& 178& 167&  27&   -1&   6&  -5&  -7&   1\\
434 &   24&  54& 218& 203&  34&   24&  53& 215& 197&  32&    0&  -1&  -3&  -6&  -2\\
433 &   16&  29&  70&  72&  23&   20&  33&  66&  73&  22&    4&   4&  -4&   1&  -1\\
432 &   12&  18&  27&  23&  16&   12&  20&  26&  27&  15&    0&   2&  -1&   4&  -1\\
\hline
\label{tabE}
\end{tabular}
\end{center}
\end{table*}

\clearpage

\end{document}